\documentstyle[floats,twocolumn,prl,aps]{revtex}

\input epsf
\input rotate

\begin{document}

\title{Self-organized Vortex State in Two-dimensional {\it Dictyostelium}
Dynamics}

\author{Wouter-Jan Rappel, Alastair Nicol, Armand Sarkissian and Herbert
Levine}
\address{Dept. of Physics \\
University of California, San Diego \\ La Jolla, CA  92093-0319}
\author {William F. Loomis}
\address{ Dept. of Biology \\
University of California, San Diego \\ La Jolla, CA  92093
}

\author{\parbox{397pt}{\vglue 0.3cm \small
We present results of experiments on the dynamics of
{\it Dictyostelium discoideum} in a novel set-up which constrains
cell motion to a plane. After aggregation,
the amoebae collect into round ''pancake" structures in which the cells rotate
around the center of the pancake. This vortex state persists
for many hours and we have explicitly verified that the motion
is not due to rotating waves of cAMP.
To provide an alternative mechanism for the self-organization of
the {\it Dictyostelium} cells, we  have developed a new model of the
dynamics of  self-propelled deformable objects.
In this model, we show that cohesive energy between the cells,
together with a coupling between the self-generated propulsive
force and the cell's configuration produces a self-organized
vortex state.
The angular velocity profiles of the experiment and of the model
are in good quantitative agreement.
The mechanism for self-organization reported here
can possibly explain similar vortex states in other biological systems.
}}

\maketitle

Spontaneous organization of self-propelled
particles can be found in a variety of systems.
Examples include the flocking of birds \cite{flocking},
the movement of traffic \cite{traffic} and
pedestrians \cite{pedes}
and the collective motion of ants \cite{ants}.
Not surprisingly, the physics of self-propelled
particles has recently attracted considerable attention \cite{work}.
The systems have been theoretically analyzed using
continuum, Navier-Stokes like, equations and discrete numerical models.
The numerical models have in common that
the objects are treated as point-particles.
However, in a variety of systems the actual shape and
plasticity of particles play an important role and hence they can not
be modeled in this oversimplified manner.
In this Letter we present experimental data
in one such system,  {\it Dictyostelium discoideum} cells, and
introduce a model that treats cells
as deformable objects rather than point particles. Our results provide
evidence for a localized vortex state in this biological system.

The developmental dynamics whereby {\it Dictyostelium discoideum} is
transformed
from a solitary amoeba state to a functional multicellular organism is
of interest to both biologists and non-equilibrium physicists~\cite{reviews}.
Past
efforts have elucidated the nonlinear chemical wave signals used to guide
aggregation~\cite{camp,darkfield,theories} as well as the chemotactic
instability
~\cite{streaming} which causes a density
collapse to one-dimensional ``streams" of incoming cells. However, much less
is known regarding subsequent events, especially with regard to organized cell
motion in the later, multicellular stages of the day-long developmental
process.

\begin{figure}[t]
\def\epsfsize#1#2{0.35#1}
\newbox\boxtmp
\hbox{\epsfbox{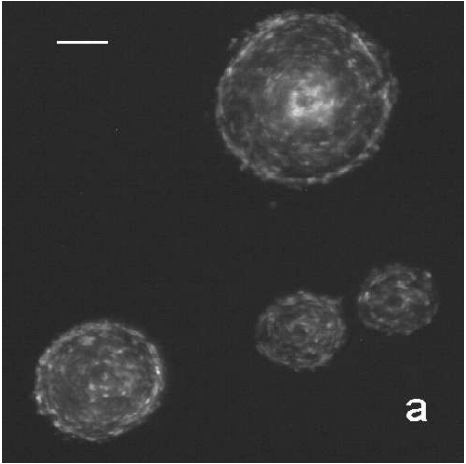}}
\vspace{.5cm}
\hbox{\epsfbox{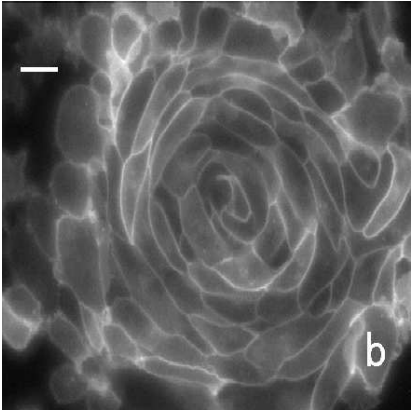}}
\vspace{0.5cm}
\caption{
Pictures of the vortex state in the experiments.
In (a) the entire cell body is fluorescent while in
(b) only the cell membrane is fluorescent.
The bar represents 100 $\mu m$ in (a) and 10 $\mu m$ in (b).
}
\label{exp}
\end{figure}
Dictyostelium cells are grown in liquid media and plated without nutrients
onto a glass surface~\cite{new}.
An additional thin layer of agarose is then overlaid on the cells. This has
the effect of restricting cell motion to the plane; in fact, the multicellular
states that form are at most a few monolayers deep. Cells aggregate normally
and form round ''pancake" structures. In Fig. 1a, we have presented a typical
snapshot of the system. In each of the observed structures,  the
cells have organized their motion into a coordinated
vortex state in which they rotate around the center of the
pancake. The rotation can be either clockwise or anti-clockwise
depending on initial conditions
and  can persist for tens of hours. Fig 1b shows a close-up
of one structure, where now the cells have been illuminated by using a
strain in which the gene for green fluorescent protein~\cite{gfp} has been
fused to
the CAR1 (cyclic AMP receptor) gene~\cite{CAR}; the expression of this gene
leads to a membrane-localized fusion protein which causes the cell to be
fluorescently outlined, as shown. This new protocol for Dictyostelium
development allows one to track cell motions in much greater detail than
has been possible to date.

It has been previously noted~\cite{rotational} that coherent rotational motion
can often be
seen in three dimensional Dictyostelium mounds, albeit as a short-lived
transient prior to cell-type sorting and tip formation at the mound apex. This
motion has been attributed~\cite{weijer} to cells moving chemotactically to
rotating
waves of cyclic AMP. To test this hypothesis, we have repeated our experiments
with a
non-signaling strain of Dictyostelium~\cite{kuspa}. Aside from the need for a
higher
initial density to overcome the inability of the system to support long-range
aggregation, the system behaves in a similar manner and produces rotating
vortex structures.
Thus, guidance via cAMP waves is {\it not} a necessary ingredient
for organized rotational motion. Instead, we suspect a self-organization of
the system similar to that seen in molecular dynamics simulations
of a confined set of particles undergoing dissipative
collisions\cite{dupetal,hem}, albeit with
deformable particles and without an artificially imposed box.

\begin{figure}[t]
\def\epsfsize#1#2{0.35#1}\newbox\boxtmp
\setbox\boxtmp=\hbox{\epsfbox{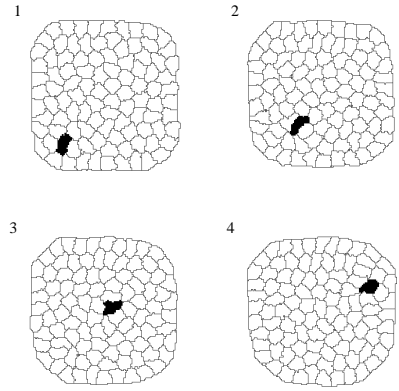}}
\rotr{\boxtmp}\vspace{0.5cm}
\caption{
Movement of a single self-propelled cell (black) surrounded
by non-propelled cells. Snapshots are taken every 800 MCS.
}
\label{single}
\end{figure}

To verify that such a self-organized state is indeed possible, we turn to
a new model of cell motion. Following Glazier and
Graner \cite{gg,yi}, we assume that cells move via (roughly)
volume-preserving fluctuations of their shape. These fluctuations are
driven by an effective free energy~\cite{details} which incorporates specific
biophysical mechanisms appropriate to Dictyostelium cells. First, cells
stick to each other and tend to move in a manner which maximizes cell-cell
boundaries over cell-medium ones~\cite{bozzaro}. In addition, each cell
contains an active cytoskeleton which can
generate forces by cycles of front protrusions and back
retractions~\cite{motion}.
This force appears in our model as a potential which preferentially accepts
fluctuations which move the cell in the direction of this propulsive force.
To qualitatively verify this methodology, we have simulated the movement of
a single self-propelled particle, as shown in Fig. \ref{single}.
The sequence clearly illustrates the deformability of the cell
and reproduces qualitatively the observed movement of a {\it Dictyostelium}
cell.

In the aforementioned previous work on flocking
and related systems, a key insight is that local interactions can
cause local velocity correlations which then causes a macroscopically
ordered state to emerge.  In our model, this suggests that one
requires that a cell adjust its own propulsive force
based on its interactions with its neighbors. We have investigated several
mechanisms whereby this could be incorporated in our model.  The most biologically
plausible assumption is that the direction of the force is updated so as to
better match the forces exerted by neighboring cells - essentially a
minimal frustration hypothesis, which is in accord with what one observes
directly in the video-microscopy. The results that follow were derived for this
specific assumption, but depend very weakly on any details beyond the basic
correlation idea.

\begin{figure}
\def\epsfsize#1#2{0.60#1}
\newbox\boxtmp
\setbox\boxtmp=\hbox{\epsfbox{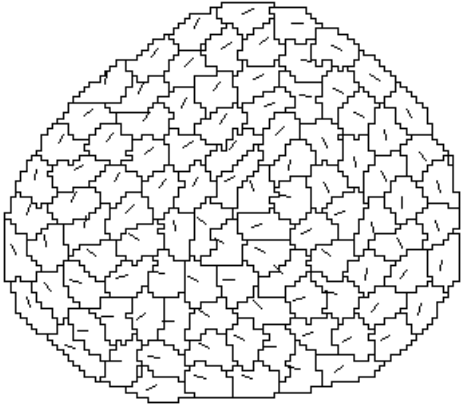}}
\rotr{\boxtmp}
\vspace{0.5cm}
\caption{
Snapshot of a typical final state of the model.
The solid lines within each cell start at the CM of the
cell and point in the direction of
the force. The parameter values for this and all other figures
in this Letter, are:
$N=100$, $A_{target}=100$, $\lambda=10$, $J_{cc}=5$, $J_{cm}=15$,
$C=1.0$ and effective ``temperature" $T=5$. The lattice contains 200x200 sites and
the force direction is updated every 2.5 MCS.
}
\label{snap}
\end{figure}

To investigate the possibility of coherent vortex motion we
started simulations with 100 square cells stacked in a
square surrounded by medium. The force direction of each cell was chosen
at random.
A snapshot of a typical final state is shown in Fig \ref{snap}
which shows the boundaries of the cells and the force direction
as a line which starts at the center-of-mass (CM) of the
cell and which points in the direction of the force.
The cells form a roughly circular patch and
are rotating around the center of the patch.
The final state is typically reached after a transient of 100-1000 Monte Carlo
steps (MCS) 
\cite{details}.
Depending on the initial conditions, the cells will rotate
either clockwise or anti-clockwise. Other than the sense of rotation and the
duration of the transient phase, nothing depends on the starting state.
Similar results were obtained for different numbers of cells. The existence
of a localized rotating state is a robust consequence of the model,
present as long as the cells are sufficiently cohesive ($J_{cm} > J_{cc}$).
Hence, we have proven that simple models based directly on the
observed cell motions can indeed account for the self-localized
vortex.

\begin{figure}[t]
\def\epsfsize#1#2{0.35#1}
\newbox\boxtmp
\setbox\boxtmp=\hbox{\epsfbox{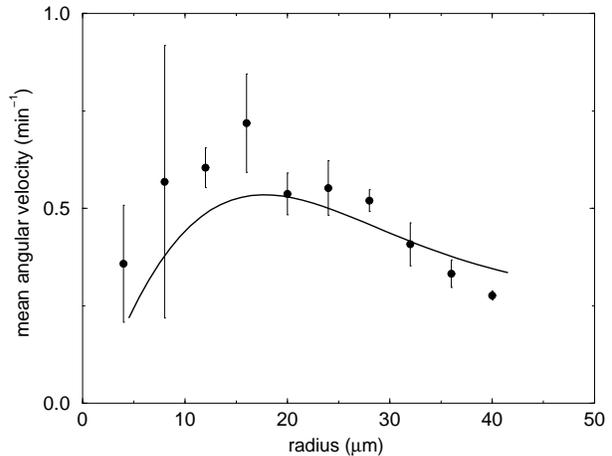}}
\rotr{\boxtmp}
\vspace{0.5cm}
\caption{
The angular velocity
as a function of the radius measured in the experiments (solid circles)
and calculated using the model (solid line).
The overall time scale in the simulations was adjusted to provide the
best fit of the model to the data.
}
\label{veloc}
\end{figure}

To further test our model, we consider the angular velocity of cells
as a function of radius. In the experiments, the angular velocity was
obtained by direct tracking of
cells. This tracking is greatly facilitated by the
use of the cell-membrane outlining approach as shown in Fig. 1b.
In detail, we have taken six separate sequences of 15 min. each and 
measured the angular velocity every 8 s.
During each sequence the radius of the individual cells changed little.
Next, we grouped the data in radius intervals of 4 $\mu m$
and calculated the average velocity and the standard deviation
for each interval. 
The data is shown in Fig. \ref{veloc} 
where the vertical bars represent one standard deviation.
The overall time scale
corresponding to a MCS was adjusted to provide the
best fit of the model (solid line) to the data; this yields 1 $MCS$ as 
$0.006 min$. As a consistency check,
we note that this gives an isolated cell velocity of 8 $\mu$m/min,
which is very close to the experimentally reported 
value of 10 $\mu$m/min \cite{canetal}. 
Although there is some way to go for
a fully quantitative theory, our simple  model does surprisingly well in
capturing both the tendency of the cell-cell interaction to speed up the
motion and the tendency of cells to slip as they move past each other,
thereby limiting the angular velocity at large radii.

What additional predictions does our model make? As the model proposes that
cell-cell adhesion is the cause of the localized coherent state, mutant
strains with reduced adhesion should not be able to form this structure. Also,
mutants with weakened cytoskeletons would not be able to organize their
motion. As chemical signaling is not necessary, disrupting the {\em external}
cAMP concentration should have minimal effects.
Finally, our model suggests that the time scale for organization should
roughly be tens of minutes for aggregates with hundreds of cells; this is
in qualitative agreement with our observations (data not shown). 

Vortex structures have been seen in other microorganism systems, namely
the nutrient-limited spreading of the newly named bacterium { \it Paenibacillus
dendritiformis}~\cite{Ben-Jacob},
and in {\it Bacillus circulans}~\cite{circulans}.
The motion of bacteria occurs through flagella
and is thus fundamentally different from  the amoeboid motion of Dictyostelium.
However, it is tempting to speculate that in these cases as well
velocity correlations induced by cell-cell interactions as well as
cohesive forces may be enough to account for these structures~\cite{chemical}.
The fact
that these disparate systems exhibit such strikingly similar non-equilibrium
structures offers comfort to the physicist proposing a simplified model for
an inscrutably complex biological process.

In summary, we have documented the existence of a localized,
coherent vortex state in Dictyostelium. Furthermore, we have
argued both experimentally and via construction of a new model that
the coherent motion is self-organized and not the result of
a rotating chemical-wave guiding the motion. The advantage of this
deformable-cell model
is that it allows for the direct comparison of simulation
with observation. In the future, we plan to extend our calculations to
the later stage processes of cell sorting and slug formation.

We would like to thank A. Kuspa and P. Devreotes for providing some of the
{\it Dictyostelium} strains used in the experiments. Also, one of us (HL)
acknowledges useful conversations with E. Ben-Jacob. This work was supported in
part by NSF DBI-95-12809.


\begin{thebibliography}{20}
\bibitem{chemical}  Ben-Jacob and collaborators have proposed in
\protect\cite{bj} that a
``rotational chemotaxis" term, arising via a correlation between the
concentration
of some chemical and the speed of the individual cell, can account for
self-localization. At least in Dictyostelium, there is no apparent need for
such a mechanism.
\end{thebibliography}
\end{document}